\documentclass[aip,rsi, amsmath,amssymb,reprint]{revtex4-1}
\usepackage[colorlinks=true,linkcolor=blue,citecolor=blue,allcolors=blue]{hyperref}
\usepackage[dvipsnames]{xcolor}
\usepackage{gensymb}
\usepackage{graphicx}
\usepackage{dcolumn}
\usepackage{multirow}
\usepackage{bm}
\usepackage{xcolor}
\definecolor{green78}{cmyk}{0.99,0,0.52,0}
\definecolor{dgreen}{rgb}{0.,0.6,0.}
\definecolor{turk}{RGB}{0,206,209}
\definecolor{kirmizi}{RGB}{128,0,0}
\definecolor{red1}{RGB}{220,20,60}

\newcommand{\mean}[1]{\left \langle #1 \right \rangle}

\begin{document}

\preprint{AIP/123-QED}

\title{Statics and Dynamics of Polymeric Droplets on Chemically Homogeneous and Heterogeneous Substrates}

\affiliation{Department of Physics, Istanbul Technical University, Maslak, Istanbul, Turkey}

\author{{\"O}. {\"O}zt{\"u}rk} 
\email{ozlemozturk@itu.edu.tr}

\author{J. Servantie}
\email{cservantie@itu.edu.tr}

\date{\today}

\begin{abstract}
  We present a molecular dynamics study of the motion of cylindrical polymer droplets on
  striped surfaces. We first consider the equilibrium properties of droplets on different
  surfaces, we show that for small stripes the Cassie-Baxter equation gives a good approximation
  of the equilibrium contact angle. As the stripe width becomes non-negligible compared to the dimension
  of the droplets, the droplet has to deform significantly to minimize its free energy, this results in a smaller
  value of the contact angle than the continuum model predicts. We then evaluate the slip length, and thus
  the damping coefficient as a function of the stripe width. For very small stripes, the heterogeneous surface
  behaves as an effective surface, with the same damping as an homogeneous surface with the same contact angle. However,
  as the stripe width increases, damping at the surface increases until reaching a plateau. Afterwards, we
  study the dynamics of droplets under a bulk force. We show that if the stripes are large enough the droplets
  are pinned until a critical acceleration. The critical acceleration increases linearly with stripe
  width. For large enough accelerations, the average velocity increases linearly with the acceleration, we show
  that it can then be predicted by a model depending only the size of droplet, viscosity and slip
  length. We show that the velocity of the droplet varies sinusoidally as a function of its position on the
  substrate. On the other hand, for accelerations just above the depinning acceleration we observe a
  characteristic stick-slip motion, with successive pinnings and depinnings.
\end{abstract}

\pacs{05.60.−k,05.70.Ln,05.70.Np,02.70.Ns,47.55.D−}
\keywords{Molecular Dynamics Simulation, Wetting on Homogeneous Substrates, Polymeric Droplets,
Equilibrium Contact Angle, Striped Substrates, A Stick-Slip Motion}
\maketitle

\section{Introduction}
The wetting behaviors of liquid volumes ranging from 
micro-/nano-liters to picoliters in microchannels are of great importance for designing
droplet-based micro-/nano-fluidic devices
\cite{SquiresTM_RMP77_2005,SeemannR2012,SchiphorstJ_LabChip18_2018}  such as
DNA-chips, \cite{DugasV_Langmuir21_2005} Lab On A CD, \cite{KongLX_JOLA2016} inkjet printing
technology, \cite{ZhouH_Langmuir33_2017} or \textit{in situ} investigation of fibrin
networks\cite{EvansHM_LOC2009}. Accordingly, equilibrium and dynamic wetting behaviors of liquid
drops on smooth (ideal, homogeneous, flat), chemically rough/structured, and topologically
patterned substrates has been studied for decades.
\cite{FurmidgeCGL_JOCS17_1962,LenzLipowsky_PRL80_1998,HGau_Science283_1999,DarhuberTroian_ARFM37_2005,ServantieMuller_JCP128_2008,JullienMC_POF2009,AguilarRL_NatMater10_2011,DupratC_Nature2012,AminiH_NatComm4_2013,YaoX_NM2013}

Furmidge\cite{FurmidgeCGL_JOCS17_1962} showed that the movement of spray liquids on different
substrates depends on the droplet's size, inclination of the substrate, the surface tension
of the drop, and the advancing and receding contact angles. Gau \textit{et al.}
\citep{HGau_Science283_1999} showed that a shape instability (a bulge state), unlike a Rayleigh 
Plateau instability, can be employed for all liquids on all striped substrates if the
hydrophilic stripes' contact angles are small enough and if these stripes are long enough.
They mentioned that these bulge states could be used to build two-dimensional microchannel
networks and hence to construct microbridges, microchips, and  microreactors.
 
To date many experimental,\cite{SchafferWong_PRL80_1998,LeopoldesBucknall_TJOPCB2005,TavanaH_Langmuir22_2006,MaheshwariS_PRL100_2008,OrejonD_Langmuir27_2011,MirsaidovUM_PNAS109_2012,Y.H.Yeong_SciRep5_2015}
computational and theoretical \cite{ShanahanMER_Langmuir11_1995,ThieleKnobloch_NJP8_2006,ThieleKnobloch_PRL97_2006,KusumaatmajaH_EL73_2006,KusumaatmajaYeomans_Langmuir23_2007,WangX.P_JOFM605_2008,QianT_JOPCM21_2009,BeltramePh_EPL86_2009,HerdeD-EL100_2012,VaragnoloS_PRL111_2013,*SbragagliaM_PRE89_2014,WangWu_SciRep5_2015,ZhangJ_Langmuir31_2015,LiuChen_PhysFluids29_2017} studies on the dynamic wetting behavior of
droplets on textured/rough surfaces have shown that a stick-slip type of microscopic slipping is
a common behavior.

Sch\"{a}ffer and Wong \cite{SchafferWong_PRL80_1998} showed that the surface roughness 
is a key factor in pinning of water in glass capillaries. L\'{e}opold\`{e}s and
Bucknall \citep{LeopoldesBucknall_TJOPCB2005} studied the spreading of droplets on chemically
heterogeneous striped substrates. In an intermediate regime, they observed a stick-slip
behavior, in other words a sudden hopping of the drop while crossing the boundary of two adjacent
stripes with different wettability. Tavana \textit{et al.}\citep{TavanaH_Langmuir22_2006} 
investigated a number of chemically different organic ($n$-alkanes) drops on two distinct
polymeric films. The observed that on the homogeneous surfaces, all of the liquids 
move smoothly, whereas on the heterogeneous polymeric films, liquids which have compounds
with short-chains present a stick-slip behavior. They noted that the cause of this stick-slip
pattern is the varying adsorption of vapor molecules on these polymer films. Maheshwari
\textit{et al.}\citep{MaheshwariS_PRL100_2008} observed  a stepwise (discontinuous)
pinning-depinning cyclic behavior and multiring stain-formations of the droplets of aqueous
DNA solutions  at high and intermediate DNA concentrations. Orejon
\textit{et al.}\citep{OrejonD_Langmuir27_2011} observed that the magnitude of the stick-slip motion
depends on the concentration of titanium dioxide nanoparticles inside
the water drops on different hydrophobic substrates. Yeong \textit{et al.}\citep{Y.H.Yeong_SciRep5_2015} 
experimentally investigated microscopic receding three-phase contact line dynamics of water droplets
on superhydrophobic substrates with regular textured pillars and with nanocomposite coatings
at micron-time/length scales. They proposed that the microscopic receding
contact line's dynamic on these  surfaces looks like `a slide-snap behavior'. 
In this motion, the receding contact line keeps on moving on the top surface
of a pillar until snapping to the adjacent pillar in contrast to a stick
and slip motion in which the microscopic receding contact lines stay
pinned (stick) before jumping into the consecutive locations.

Numerical and/or theoretical investigations into these dynamic behaviors can also be summarized
as the following. Shanahan \citep{ShanahanMER_Langmuir11_1995} calculated the excess free energy
per unit length of the three-phase contact line of a spherical evaporating droplet on an ideal
solid surface by taking into account the solid-liquid, solid-vapor, and liquid-vapor interfacial
free energies of this system. The system must overcome this potential energy barrier
so that the triple line can jump to its next static anchored position after this pinning state
in which case the contact radius of the drop remains constant while the contact angle and the
volume of the droplet decreases. Thiele and Knobloch,\citep{ThieleKnobloch_PRL97_2006, ThieleKnobloch_NJP8_2006} and Beltrame \textit{et al.} \citep{BeltramePh_EPL86_2009} modeled pinning/depinning
cycle of moving two- and/or three-dimensional mesoscopic droplets with a wetting layer on
heterogeneous surfaces by solving the Navier-Stokes equation within the lubrication
approximation. \citep{OronA_ROMP69_1997} Kusumaatmaja and
co-workers \citep{KusumaatmajaH_EL73_2006,KusumaatmajaYeomans_Langmuir23_2007} described
equilibrium  properties of liquid droplets on substrates with different wettability strengths
thanks to the Landau free energy. \citep{BriantAJ_PRE69_2004_I,*BriantAJYeomansJM_PRE69_2004_II}
The free energy they choose causes a liquid droplet to coexist with its vapor phase. They computed
the dynamic properties of the liquid drops by solving the continuity and the Navier-Stokes
hydrodynamic equations of motion with free-energy lattice Boltzmann (LB) numerical
simulation technique. \citep{BriantAJ_PRE69_2004_I,*BriantAJYeomansJM_PRE69_2004_II} Wang
\textit{et al.} \citep{WangX.P_JOFM605_2008} carried on continuum simulations of contact
line dynamics of a binary fluid flowing between two chemically patterned solid substrates
using a diffuse-interface model with the generalized Navier boundary condition. In this model
they solved numerically two coupled equations of motion which consist of the convection-diffusion
equation for a field variable and the Navier-Stokes equation with a capillary force density. Their
simulation results showed an oscillatory (stick-slip) behavior of the interface of the two immiscible
fluids on the substrate. Qian \textit{et al.} \citep{QianT_JOPCM21_2009} used both Molecular Dynamics
simulations (MD) and a continuum model to investigate the nanoscale-hydrodynamics of the moving
contact line on chemically striped surfaces. They observed a stick-slip motion of the contact
line on these substrates. Herde \textit{et al.} \citep{HerdeD-EL100_2012} studied the
depinning/repinning dynamics of two-dimensional drops driven by a lateral body force on chemically
heterogeneous flat surfaces having periodic (sinusoidal) wetting energy. They solved the Navier
Stokes equation for small Reynolds numbers numerically using a boundary element method. Sbragaglia
and co-workers \citep{VaragnoloS_PRL111_2013,*SbragagliaM_PRE89_2014} observed a stick-slip
periodic behavior of
liquid drops sliding over solid substrates patterned with parallel stripes of varying wettability
degrees both experimentally and numerically in two dimensions. They solved the diffuse-interface 
Navier-Stokes equations of motion for a binary mixture using LB simulation method. Wang and Wu
\citep{WangWu_SciRep5_2015} investigated the stick-slip motion of moving contact line of the evaporating
liquid drops on solid substrates with flexible nano-pillars thanks to MD simulations. Zhang
\textit{et al.} \citep{ZhangJ_Langmuir31_2015} studied the evaporation of liquid cylindrical drops
on chemically heterogeneous surfaces with alternating stripes of two types of equal widths also
with MD simulations. They found that at the microscopic scales, the three-phase contact line
is moving slowly instead of pinning (stick) entirely  at the boundary between the two distinct
stripes with the different attraction or energy parameters and the roughly $48\degree$ wettability
contrast.

This paper is organized as follows: In Sec. II,  we outline
the details of the coarse grained model and the MD simulation
technique that we use in this work. Then, in Sec. III  we study the 
equilibrium and dynamic properties of polymeric droplets on homogeneous and
heterogeneous surfaces. The conclusions are finally drawn in Sec. IV.

\section{The Coarse-Grained MD model} 
In this paper, we use a generic particle based molecular dynamics (MD) simulation
technique to study the static and dynamic wetting behaviors of polymer droplets on
different substrates. \cite{GrestKremer_PRA33_1986,KremerGrest_JCP92_1990,BennemannC}
Thanks to this coarse-grained model one can investigate the universal wetting
properties of polymeric droplets on corrugated or smooth substrates. In this
coarse-grained model, a bead of a linear homopolymer chain actually corresponds
to a group of united molecules/atoms. The advantage of using polymer melts in
MD simulations is due to the fact that their vapor pressure is very low.
\cite{ ServantieMuller_JCP128_2008, ServantieMuller_PRL101_2008} Hence, the number
of atoms in the vapor phase remains small permitting the study of larger systems.

The polymer melt is modeled with bonded (intramolecular) and nonbonded (intermolecular)
interactions. The bonded interactions are between neighboring beads of a polymer,
it is modeled by the finitely extensible nonlinear elastic (FENE) potential,
\cite{GrestKremer_PRA33_1986,KremerGrest_JCP92_1990}
\begin{equation}
{U}_{\rm FENE}=\left\lbrace
\begin{array}{cc}
-  \frac{1}{2} k {R_{0}}^{2}  \ln  \left[ 1  -  \left  
(  \frac{r}{R_{0} } \right )^2  \right ]  &  \text{for} \; r< R_{0}   \\ 
\infty  & \text{for} \;r \ge R_{0} 
\end{array}
\right.
\label{eq:fene}
\end{equation}
where the spring constant is $k=30 \epsilon/\sigma^{2}$ and the maximum covalent bond
length $ R_{0}=1.5 \sigma $. Thanks to the FENE potential, the connectivity of the
beads along the backbone chain is obtained. In addition to the bonded potential, there
is a 12-6 Lennard-Jones (LJ) potential between each pair of beads in the system,
\begin{equation}
{U}_{\rm LJ}=\left \lbrace
\begin{array}{cc}
  4 \epsilon \left [  \left ( \frac{ \sigma }{r} \right )^{12}
    -  \left ( \frac{ \sigma }{r} \right )^6  \right ]
& \text{for} \;  r < r_{c}    \\
0  & \text{for} \; r  \ge  r_{c} 
\end{array}
\right.
 \label{eq:LJ}
\end{equation}
where the cut-off distance  is $r_{c}=2  \times 2^{1/6 } \sigma$. The repulsive part
of the Lennard-Jones interaction permits to enforce the excluded volume effects
while the attractive part permits to have a liquid state. The system is prepared so
that each polymer contains $N_{p}=10$ identical monomeric beads of mass $m$.

The surface is modelled by two rigid layers of face-centered-cubic lattice. The number
density of the substrate is chosen as $\rho_{s} =2.0 \sigma^{-3} $.
\cite{ServantieMuller_JCP128_2008} Large enough so that no polymer atoms go through the surface. The
atoms of the substrate interact with the polymeric fluid with a modified
Lennard-Jones potential, \cite{BarratBocquet_FaradayDiscuss112_1999}
\begin{equation}
U_{s}=\left\lbrace
\begin{array}{cc}
4 \epsilon_{s}  \left [  \left ( \frac{ \sigma_{s} }{r} \right )^{12} 
- C_{s}  \left ( \frac{ \sigma_{s}  }{r} \right )^6  \right ]
&   \text{for} \; r < r_{c}    \\  
0  &  \text{for} \; r  \ge  r_{c} 
\end{array}
\right.
\label{eq:LJs}
\end{equation}
where the cut-off distance is the same as in Eq. (\ref{eq:LJ}). 
The length and energy scales of the potential energy are fixed to
$\sigma_{s}=\sigma$ and $\epsilon_{s}=\epsilon$, respectively. Finally, the
empirical parameter $C_{s}$ quantifies the hydrophobicity of the surface.
The larger $C_{s}$ is, the more attractive and consequently, hydrophilic
the substrate is. Thus, one can easily tune the wetting properties of the
surface. Furthermore, one can construct a simple heterogeneous surface by
alternating the type of atom by using different hydrophobicity
parameter $C_s$. We prepare surfaces with increasing stripe width. The hydrophobicity
parameters are fixed to $C_{s}=0.4$ (hydrophobic surface) and $C_{s}=0.6$ (hydrophilic
surface). All the surfaces have a total of $11520$ atoms and the dimensions
$L_x=241.90489$ (longitudinal to the flow), $L_y=18.9\sigma$, and
$L_z=150.0\sigma$ (transverse to the flow). Periodic boundaries are enforced in the
$x$ and $y$ directions, while reflective periodic boundaries are in effect on the top
of the simulation box in the $z$ direction. The height of the simulation box along
the $z$ axis is taken large enough for the droplet's upper part not to touch the box,
therefore one can obtain a free liquid surface.

\begin{figure}[h!]
  \includegraphics[width=8cm,keepaspectratio=true]{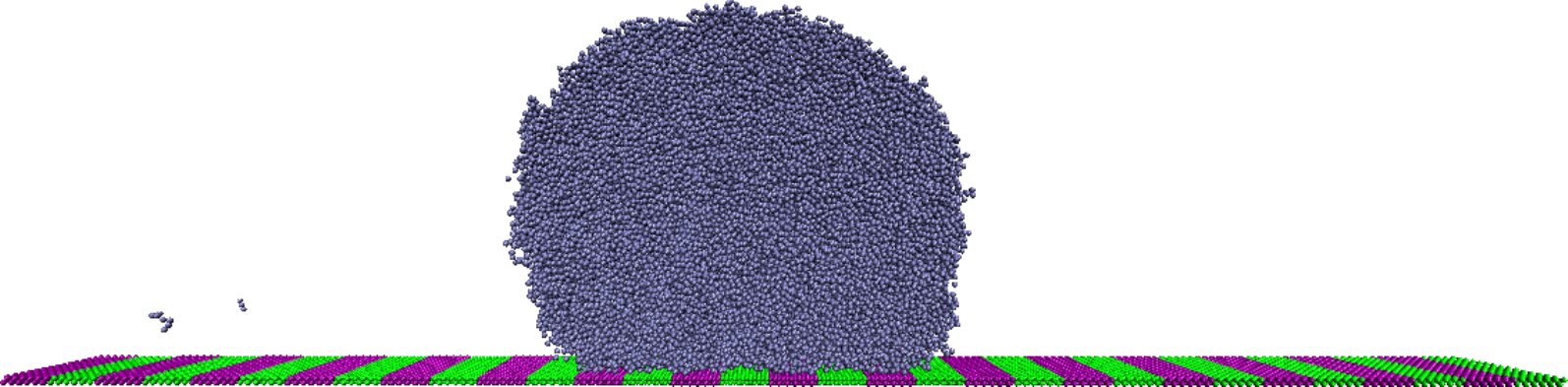} 
  \caption{\label{fig:w756eq}
  (Color Online)  MD simulation snapshot of a static droplet at equilibrium. 
The droplet with $N=50000$ segments (iceblue) is on a striped surface having
two types of atoms with $C_{s}=0.4$ (green) and $C_{s}=0.6$ (purple). The stripes
have and equal widths of $w=7.56 \sigma$.}
\end{figure}

Finally, the equations of motion are integrated with the velocity
Verlet algorithm \cite{Verlet_PhysRev159_1967} with a time step
of $\Delta t=0.005\tau$. We fix the temperature of the system to
$k_{B}T=1.2\epsilon$ for all the MD simulations, at this temperature the density
of the polymer melt is $\rho_{p} =0.788 \sigma^{-3}$ and the vapor density is
negligible. \cite{PastorinoC_PRE76_2007,GonzalezMacDowel_JCP113_2000}
Consequently, we deal with so-called "dry wetting" in this study, the surface pressure 
of the liquid polymer vapor on the substrate is extremely low. We depict
in Fig.\ref{fig:w756eq} an equilibrated droplet of $N=50000$ monomers
on a heterogeneous surface with stripe width $w=7.56 \sigma$.
The equilibrium contact angle of the droplet is $\theta_{E}=133\degree$.

A dissipative particle dynamics
(DPD)\cite{HoogerbruggeKoelman_EL19_1992,EspanolWarren_EL30_1995,ServantieMuller_JCP128_2008}
thermostat is used to keep the temperature of the system constant. The DPD
thermostat has the advantage of conserving the momentum locally instead of globally
as for the Nos\'{e}-Hoover thermostat.
\cite{HoogerbruggeKoelman_EL19_1992,EspanolWarren_EL30_1995,SoddemannT_PRE68_2003}
The damping coefficient of the thermostat is set to $\gamma=0.5 \tau^{-1}$ in
all our simulations.

\section{Results}

 \subsection{Equilibrium wetting properties}

\begin{figure}
  \includegraphics[scale=0.5]{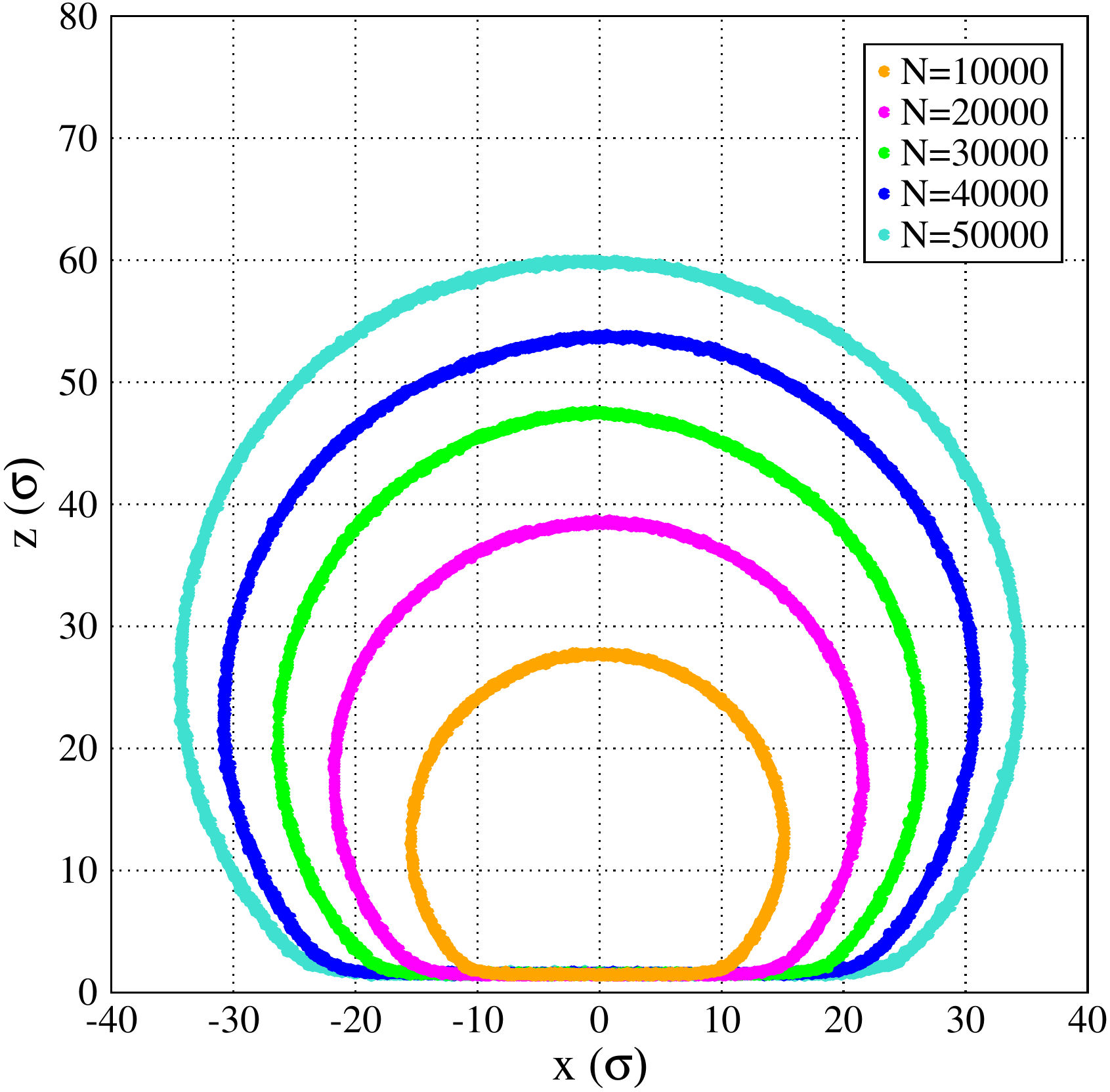}
  \caption{\label{fig:Cs05} Density profiles for increasing
    monomer sizes $N=10000,\; 20000,\; 30000, \; 40000,\; 50000$. The surface
    is homogeneous with hydrophobic parameter $C_{s}=0.5$. The averaged
    equilibrium contact angle of these droplets is $\theta_{E}=137\degree\pm1\degree$.}
\end{figure}

\begin{figure}
  \includegraphics[width=8cm]{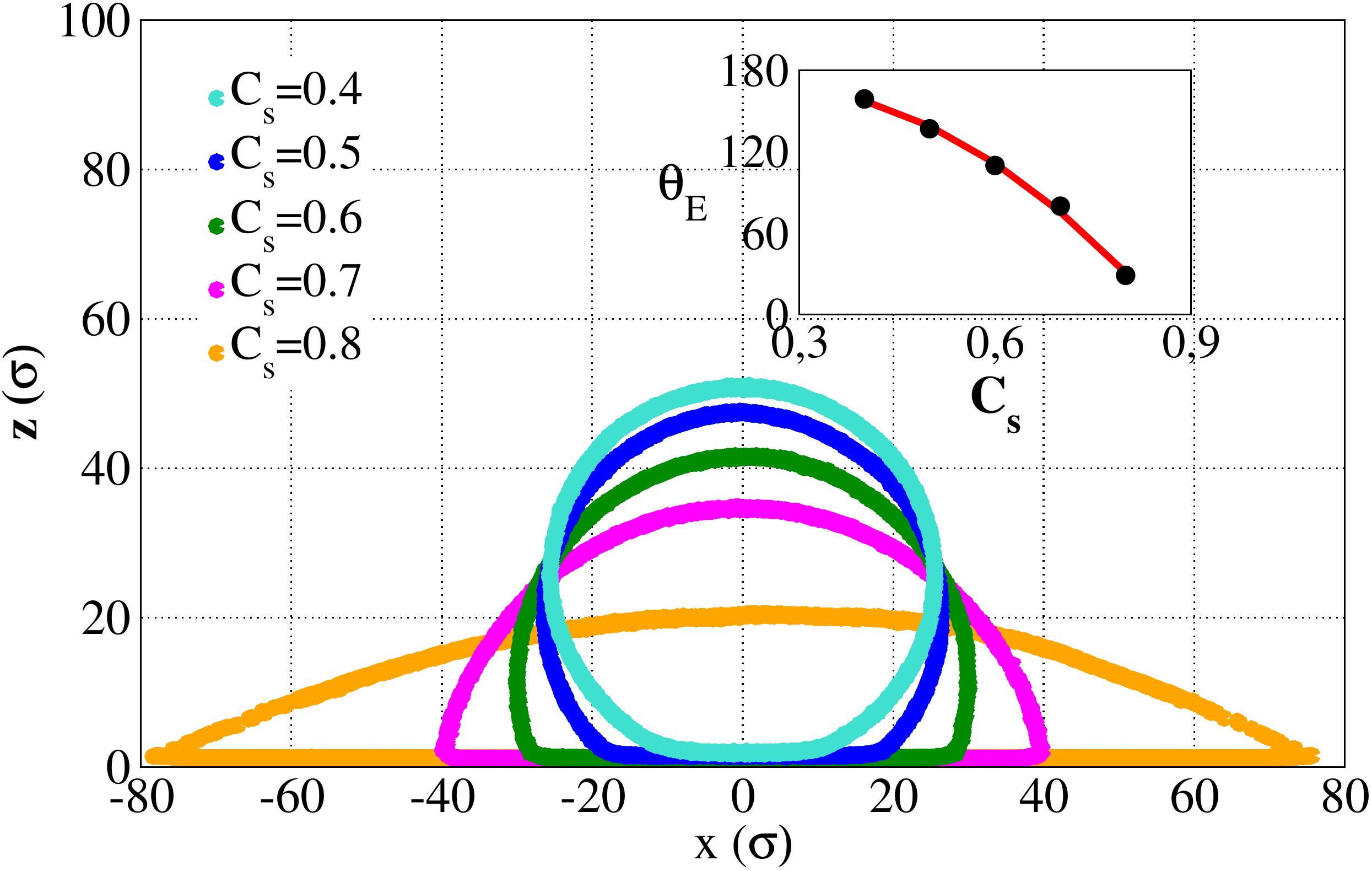}
  \caption{\label{fig:CsN30000} (Color online) Density profiles of droplets with $N=30000$
    atoms for increasing hydrophilic parameters $C_{s}=0.4,
    \;0.5,\;0.6,\;0.7,\;0.8$. The surfaces are chemically
    homogeneous substrates. The inset represents the equilibrium contact angle as a
    function of the substrate strength.}
\end{figure}

\begin{figure}
  \includegraphics[width=8cm]{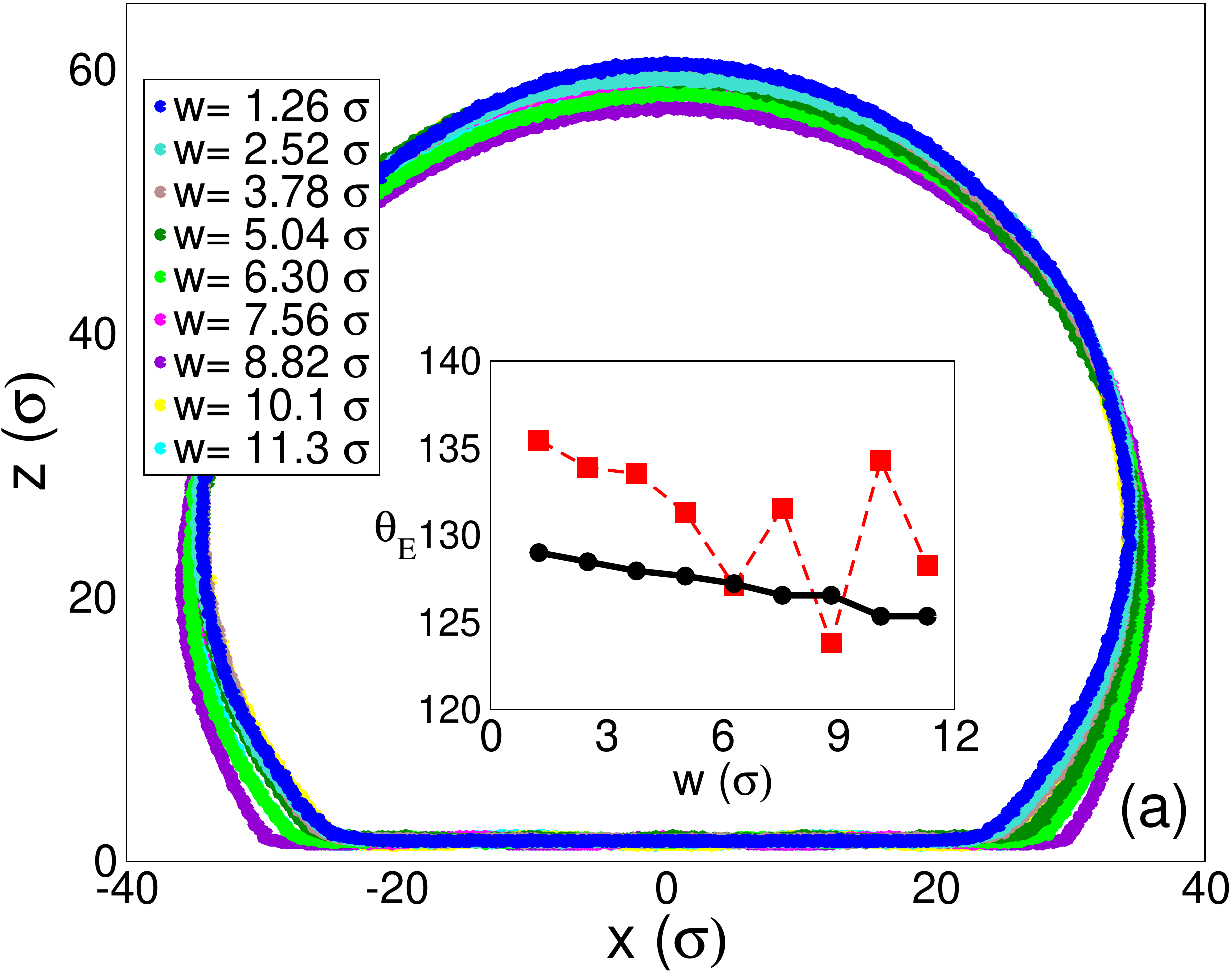}  
  \includegraphics[width=8cm]{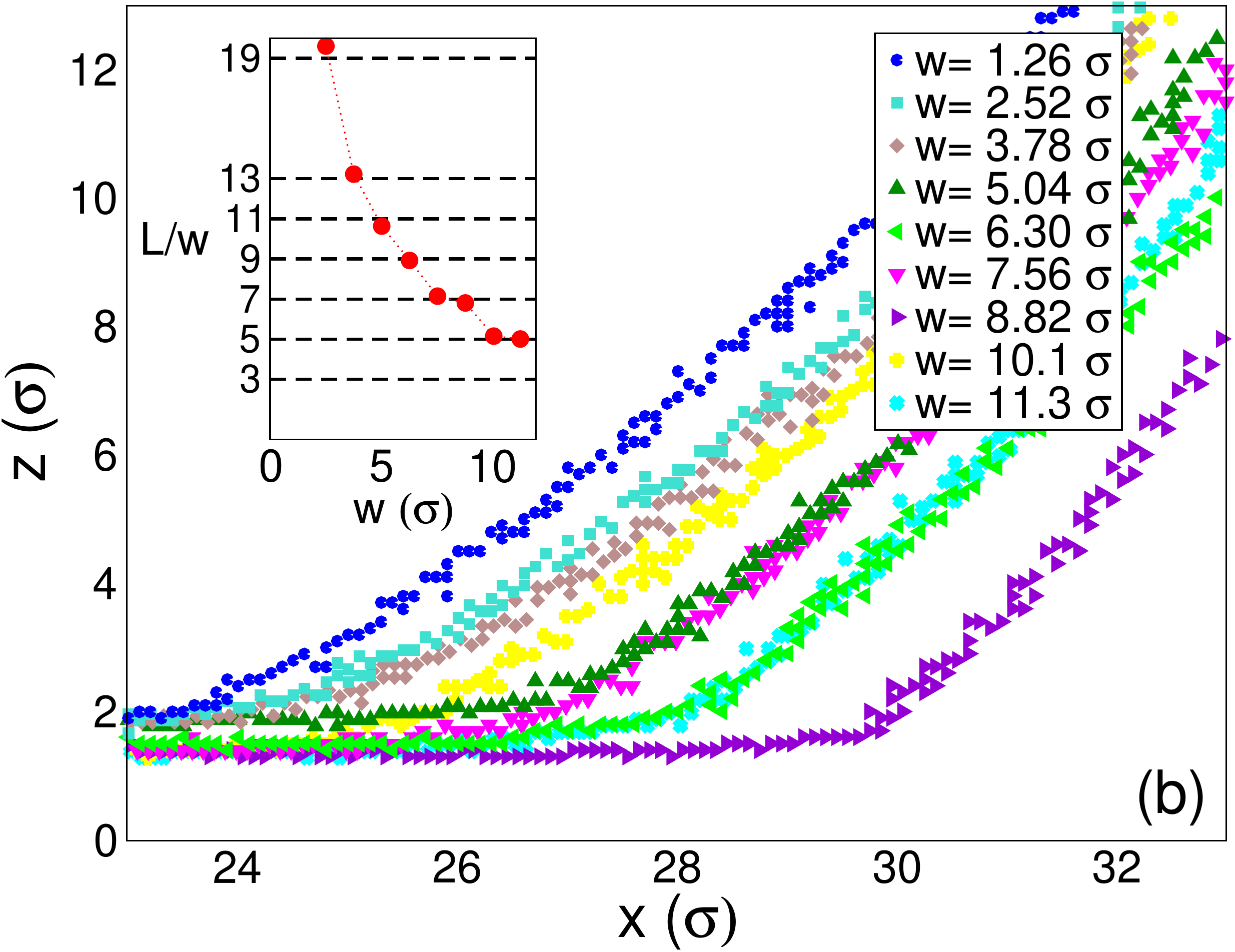}
  \caption{\label{fig:4Width50000nsivi} (Color online) (a) Density contours of a droplet
    of $N=50000$ atoms on different striped substrates. The circles in the inset represent the
    Cassie-Baxter angle, and the squares the actual equilibrium contact angle as a function
    of the stripe width $w$. (b) Close up view of the contact line for the different droplets, the inset
  depicts the length of the droplet divided by the stripe width $w$ as a function of $w$.}
\end{figure}

 We calculate the equilibrium contact angles, $\theta_{E}$, of droplets for
 various strengths of the hydrophobicity parameter, $C_{s}$,
 and droplet sizes, $N$. We focus on cylindrical droplets in order to study
 larger liquid systems and have better statistics. We use droplet sizes from
 $N=10000$ to $N=50000$ monomers and hydrophobicity parameters in the range
 $C_{s}=0.3-0.8$ to study the equilibrium density profiles. The density profiles
 are obtained by counting the number of monomers in two-dimensional boxes of size
 $0.1\sigma$ in the $x$ and $z$ directions.  We choose the contour line as the arithmetic
 mean of the densities of the polymer melt
 and its vapor, since the density of the vapor is negligible, the contour line corresponds
 to a density of $\rho_{p}/2 =0.394 \sigma^{-3}$. We depict in Fig.\ref{fig:Cs05} the
 density profiles for increasing number of monomers with the hydrophilic parameter set
 to $C_s=0.5$.
The contact angle is independent of the size of the droplet,
 we can hence evaluate the angle precisely by
 using drops of different sizes. \cite{ServantieMuller_JCP128_2008} The geometry
 of cylindrical droplets at equilibrium satisfy the following relationships, 
\begin{eqnarray}
  V&=&\frac {R^2}{2}\left( 2\theta_{E}-\sin2\theta_{E}\right) L_{y} \\
  A&=&2R\sin\theta_{E}Ly \\
  H&=&R(1-\cos \theta_E) \\
  r_{z}&=&R \left( \frac{4}{3} \frac{\sin^3\theta_{E}}{2\theta_{E}-\sin2\theta_{E}}-\cos\theta_{E}\right)
  {\label{eq:theta}}
\end{eqnarray}      
where $V$, $H$, $A$, and $r_{z}$ are respectively the volume, the height, the
area of contact with the substrate, and height of the center of mass. One can easily
find the heights, $r_{z}$. Then, using the contour plots in Fig.\ref{fig:Cs05} we fit
circles to the droplets to find the radius of the droplet $R$. 
One can then solve  Eq. (\ref{eq:theta}) numerically for $\theta_{E}$. The same
procedure is applied for varying hydrophilic parameters. We depict in
Fig. \ref{fig:CsN30000} the density  profile of a droplet of $N=30000$ monomers for
increasing values of $C_s$. As expected, increasing the strength of the attractive part
of the interaction potential results in a more hydrophilic surface, and thus a lower contact
angle. The inset of Fig.\ref{fig:CsN30000} depicts the equilibrium contact
angle as a function of $C_s$.
We now focus on the striped surfaces. We determine from the inset of Fig. \ref{fig:CsN30000}
that $C_{s}=0.4$ ($\theta_E=159\degree$) corresponds to a super-hydrophobic surface while
$C_{s}=0.6$ ($\theta_E=110\degree$) is more hydrophilic. We choose those two values for the
striped surfaces. We compute the equilibrium density profile of a droplet of $N=50000$ atoms
on surfaces of varying stripe width from $ w=1.26,\;2.52,\;
3.78,\;5.04,\;6.30,\;7.56,\;
8.82,\;10.08,\; 11.34\sigma$. These values
correspond to the stripe width to droplet length ratios of $w_{p}=3,\;5,\;8,\;10,
\;13,\;15,\;18,\;20 ,\;23\%$.
We depict in Fig. \ref{fig:4Width50000nsivi} the density profiles. 
The inset of Fig. \ref{fig:4Width50000nsivi} (a) represents the equilibrium contact
angles, and the contact angle corresponding to the Cassie-Baxter equation as
a function of the stripe width. The Cassie-Baxter equation, \cite{CassieABDBaxterS_TOTFS1944}
is a continuum result, it predicts the equilibrium contact angle of a droplet on a mixed surface as,
\begin{equation}
  \cos\theta_{CB}=f_{1}\cos\theta_{1}+f_{2}\cos\theta_{2},
\end{equation}
where $f_1$ and $f_2$ are respectively the contact area fractions of the surface of type 1 and 2, 
and $\theta_1$ and $\theta_2$ their respective equilibrium contact angles. We notice that, the
calculated value $\theta_E$ is close to the continuum prediction $\theta_{CB}$. Indeed, on average
we obtain $ \theta_{E}=131\degree\pm 4\degree$ for the equilibrium contact angle, while the data
for the homogeneous surfaces and $f_{1}=f_{2}=0.5$ yields $\theta_{CB}=127\pm 1\degree$. On the other
hand, we notice that the contact angle decreases with the stripe width and varies wildly at large values
of $w$. This is due to the fact that at equilibrium the droplet maximizes its contact area with the hydrophilic
stripes in order to minimize the free energy. In order to achieve this the droplets slightly deforms to be in contact
with one more hydrophilic stripe than hydrophobic stripes. In that case, the droplet has to be in contact with an
odd number of stripes. As the stripe width increases, in order to accommodate an odd number of stripes, the droplet has to
deform significantly, resulting in the important variation in the contact angle. We have depicted in
Fig. \ref{fig:4Width50000nsivi}(b) a close up on the contact line for all the different striped surfaces. In the inset
we give the contact length to stripe width ratio. As we see, the length of the droplet varies as an odd number times
the stripe width. In general, the fact that there is an extra hydrophilic stripe with respect to the hydrophobic one will
make the surface effectively more hydrophilic, and hence with a lower contact angle. As the stripe width increases, the 
effect becomes more important, thus the contact angle will decrease with $w$. We remark that we have taken this effect
into account while evaluating the Cassie-Baxter contact angle in Fig. \ref{fig:4Width50000nsivi}(a).

\subsection{Boundary condition}

\begin{figure}[h!]
  \includegraphics[width=8cm,keepaspectratio=true]{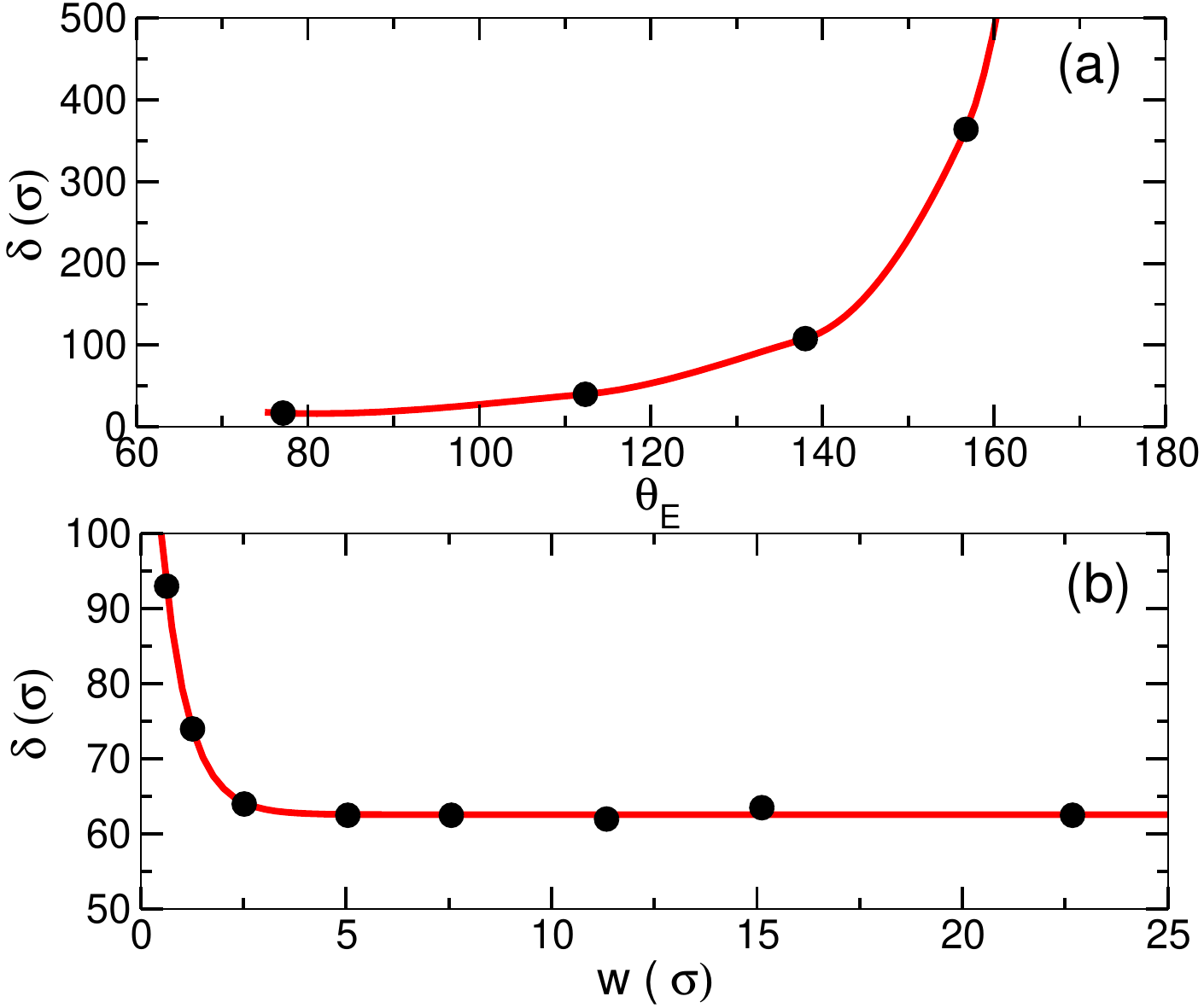}  
  \caption{\label{fig:sliplength}
    (Color online) (a) Slip length as a function of the equilibrium contact angle for
    homogeneous surfaces (b) Slip length as a function of the stripe width. The solid lines
    are guides for the eyes.}
\end{figure}

Before considering the dynamics of droplets on the different surfaces one has to evaluate the boundary
condition. Indeed, we showed previously that for microscopic droplets the presence of
slip at the surface can significantly affect the dynamics of the droplets. Specifically, for small droplets
and large contact angles the slipping on the surface becomes the dominating dissipation mechanism leading
to an increased velocity. \cite{ServantieMuller_JCP128_2008}
In the presence of slippage at the boundary one can use the Navier slip boundary condition,\cite{Navier1823} 
\begin{equation}
\eta \frac{\partial v_x}{\partial z} \Big|_{z_b}=\lambda v_b
\end{equation}
where $z_b$ is the position of the boundary, $\lambda$ a damping coefficient quantifying the friction
at the surface, and $v_b$ the velocity of the fluid on the surface. The damping coefficient can be
evaluated thanks to a Green-Kubo relationship, namely,\cite{BocquetBarrat_PRL_70_1993,BocquetBarrat_PRE_49_1994}
\begin{equation}
  \lambda=\frac{1}{k_B T A} \int_0^\infty dt\ \mean{F_s(t) F_s(0)}
  \label{eqlambda}
\end{equation}
where $F_s$ is the tangential force exerted by the substrate on the fluid, and $A$ the area of
contact. The slip length is then given as,
\begin{equation}
  \delta=\frac{\eta}{\lambda}
  \label{eqdelta}
\end{equation}
In order to evaluate $\delta$ we confine a fluid with $N=50000$ monomers between two surfaces
with $L_x=90.7\sigma$ and $L_y=18.9\sigma$. The distance between the surfaces is tuned in order
to recover the bulk liquid density far from the walls. After an equilibration, we compute the
total transverse force for $4\times 10^6$ time steps and evaluate its time
auto-correlation. Finally the slip length is obtained thanks to Eqs. \ref{eqlambda}
and \ref{eqdelta}.
We depict the results in Fig. \ref{fig:sliplength}.
For homogeneous surfaces we notice a sharp increase of $\delta$ as a function of the equilibrium
contact angle $\theta_E$. In each case, the slip length is important compared to the dimensions of the
droplet, consequently we expect slippage to be the dominating mechanism of dissipation on the
homogeneous surfaces. On the other hand, for the heterogeneous surfaces for very small widths we recover
the value of $\delta$ corresponding to the homogeneous case i.e. for a contact angle of
$\theta\approx 130\degree$ about $\delta \approx 90\ \sigma$, as the stripe width increases the slip length
decreases and rapidly reaches a plateau at $\delta \approx 60\ \sigma$. The plateau is reached at approximately
$3.5\ \sigma$, close to the effective size of the polymers. Indeed, the end-to-end distance of the polymers
is found to be $R_{\rm ee}=3.447\ \sigma$. For very small stripe widths the
stripes merge to an effective surface with an equilibrium contact angle corresponding to the Cassie-Baxter
relation, as the stripe width increase the polymers interacts with the two distinct surfaces, leading to increased
fluctuations at their boundaries, and consequently increased damping $\lambda$, therefore decreased slip length
$\delta$. Once the stripes becomes larger than the polymers, the increased damping remains confined to the boundary
between stripes, and thus a plateau is reached. We note that the slip length is important for both the homogeneous
and heterogeneous surfaces, this is due to the fact that even though the striped surfaces are chemically
heterogeneous they are still very smooth.

\subsection{Dynamic wetting properties}

\begin{figure}[h!]
  \includegraphics[width=8.5cm]{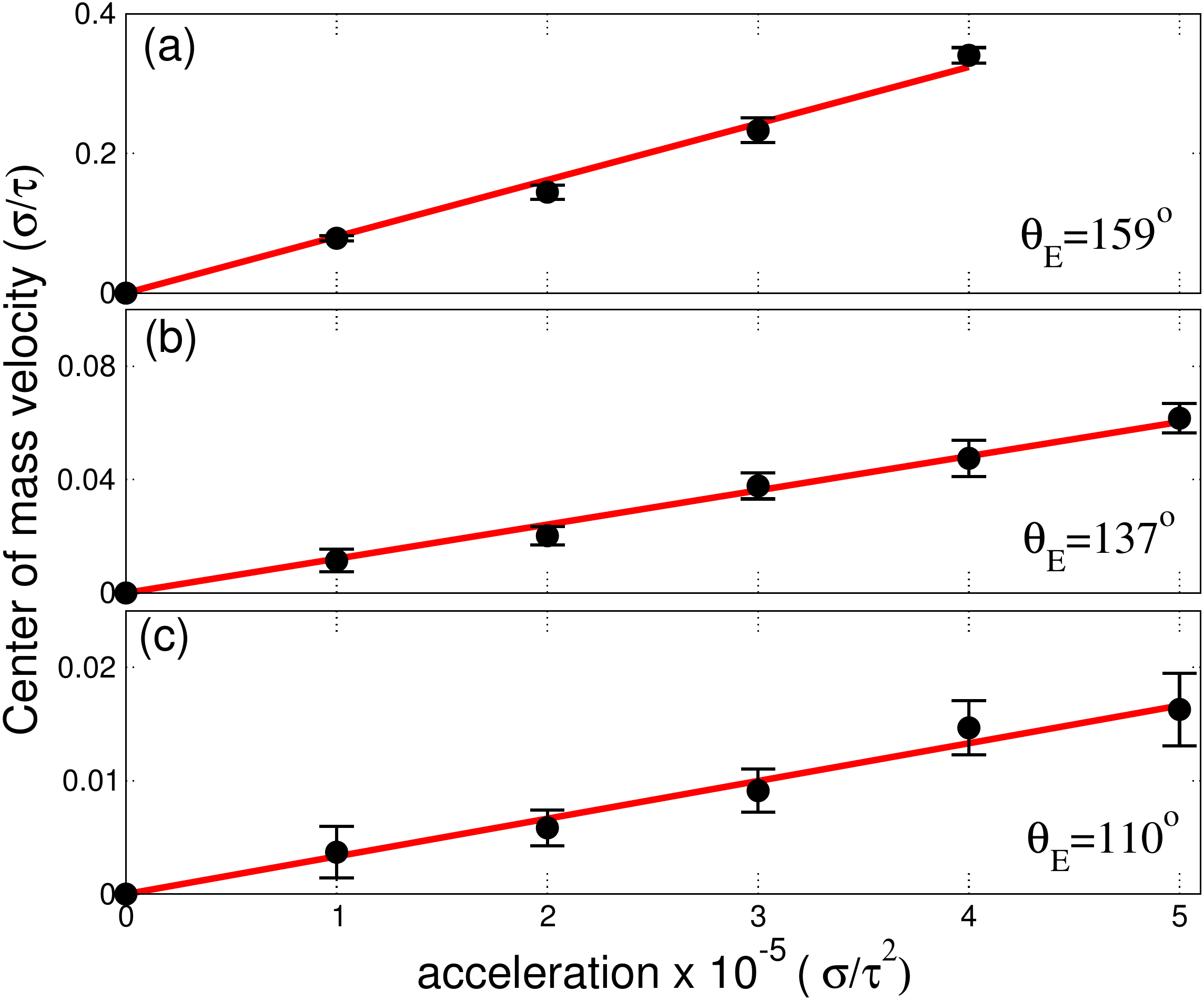} 
     \caption{\label{fig:HOMOva} (Color online)  x-component of 
the time-averaged velocity of the center of mass of the droplets for  $N=50000$ monomers as a function of 
the acceleration for the chemically homogeneous
  surfaces with increasing  wettability degrees, $(\text{\bf a})C_{s}=0.4 , \;\theta_{E}=159 \degree$,
   $(\text{\bf b})C_{s}=0.5, \; \theta_{E}=137 \degree$,
   $(\text{\bf c})C_{s}=0.6 , \;\theta_{E}=110 \degree$. The circles are the results of the MD simulations and
  the plain lines are linear fits.  The error bars correspond to the standard deviations of
  the means.}
\end{figure}

In this section we focus on the dynamics of the cylindrical polymer droplets
on the homogeneous and heterogeneous substrates. We first study homogeneous
surfaces with only one type of atom. We consider three different hydrophilicity
parameter $C_{s}=0.4,\;0.5,\;0.6$ which corresponds to the equilibrium contact angles
$\theta_E=159\degree, \;137\degree\;,110\degree$. In order a to have a sustained motion we apply
a bulk acceleration $a$ in the longitudinal direction $x$ to all the fluid atoms. The calculations
are carried out for an equilibrated droplet of $N=50000$ monomers. We use five different values
for the bulk acceleration, namely $a=0.00001,\; 0.00002,\; 0.00003,\; 0.00004,\; 0.00005$. After $2\times 10^6$
of equilibration steps, a non-equilibrium steady state is reached, we then compute the time average of the velocity of the
center mass in the longitudinal direction, $\mean{v_{\rm CM}}$ for a further $2\times 10^6$ time steps. We depict the results in
Fig.\ref{fig:HOMOva}.
We showed previously \cite{ServantieMuller_JCP128_2008} that the velocity profile in cylindrical droplets can be
estimated by
\begin{equation}
  v_x(z)=\frac{\rho_p}{\eta} \left[\left(H-\frac{z}{2}\right) z+\delta z \right]\  a
  \label{eqmodel0}
\end{equation}
thus, the velocity of the center of mass can then be written as,
\begin{equation}
  v_{\rm CM}=\frac{\rho_p}{\eta} \left[\left(H-\frac{r_z}{2}\right) r_z+\delta H \right]\  a
  \label{eqmodel1}
\end{equation}
where $H$ is the height of the droplet and $\delta$ the slip length. This relationship is valid as
long as the droplet is not deformed i.e. at small velocities. We hence expect a linear increase of the
velocity with the acceleration. One could hence model the velocity of the center of mass as,
\begin{equation}
  m_{\rm CM}\ \frac{dv_{\rm CM}}{dt}=-\gamma\ v_{\rm CM}+m_{\rm CM}\ a
  \label{eqmodel2}
\end{equation}
where the effective damping coefficient $\gamma$ comprises all the different types of dissipation
present in the droplet, namely, viscous dissipation in the volume, frictional dissipation at the
surface, and dissipation at the contact line. At the steady state the expectation value of the
center of mass velocity can then be written as,
\begin{equation}
  v_{\rm CM}=\frac{\rho_p V}{\gamma} a.
  \label{eqmodel3}
\end{equation}
Performing linear fits on the data in Fig. \ref{fig:HOMOva} permits to evaluate the effective dissipation
coefficient $\gamma$. We present the results of the fits, and the expected value of the damping coefficient
according to Eq. \ref{eqmodel1} in Table \ref{tab:HomoDynamic}.
\begin{table}[h!]
\caption{\label{tab:HomoDynamic}Molecular Dynamics simulation 
results for polymer drops of $N=50000$ atoms on chemically 
homogeneous surfaces for the three wettability degrees, $C_{s}$.
$\gamma$ corresponds to the calculated effective damping coefficient and $\gamma_{\rm MD}$ the one actually
measured during the simulation. }
\begin{ruledtabular}
  \begin{tabular}{cccccccc}
    $C_{s}$  & $\theta_E$ &  $r_z$  ($\sigma)$&  $H$ ($\sigma)$&  $\delta$ ($\sigma)$ &  $\gamma$  \; $(m/\tau)$&   $\gamma_{\rm MD}$  \; $(m/\tau)$\\  \hline
    0.4     & 159\degree  &  31   &      63   &      364  &        14   &        6 \\
    0.5    &  137\degree &  27  &       58   &      108   &       45   &        41 \\
    0.6    &  110\degree &23  &       52   &      40    &       112  &        150 \\
  \end{tabular}
\end{ruledtabular}
\end{table}
We notice that apart for $C_s=0.4$ Eq. \ref{eqmodel1} gives a relatively good approximation of the damping
coefficient. Errors come from two different approximation; firstly Eq. \ref{eqmodel1} is valid only when
the contact angle is not too large, indeed it was derived according to the lubrication approximation, secondly,
in order to evaluate the slip length with Eq. \ref{eqdelta} one needs the local viscosity i.e. the viscosity
at the surface. It is known that the mobility of the fluid atoms is affected by the strength of the substrate 
and consequently the viscosity. \cite{ServantieMuller_PRL101_2008}

We previously looked to the size dependence of the steady state velocity. \cite{ServantieMuller_JCP128_2008}
For small droplets the dominating dissipation mechanism is the friction at the surface, in that
case $v_{\rm CM} \sim R$. On the other hand, for large droplets the dominating dissipation mechanism is viscous
dissipation in the volume, then $v_{\rm CM} \sim R^2$. In general, the velocity increases with
size. Unfortunately, one can not write a simple scaling law for the dependence on contact angle in
Eq. \ref{eqmodel1}. Instead, one can look to the velocity at the top of the droplet $z=H$,
\begin{equation}
  v_{\rm top}=\frac{\rho_p a}{\eta}\ \left(\frac{H}{2}+\delta\right) H 
\end{equation}
For droplets of fixed volume one can then get two limiting cases. For small droplets or a large
slip length compared to its height the velocity at the top scales as $v_{\rm top} \sim \delta H$. While
for large droplets or a small slip length compared its size one has $v_{\rm top} \sim H^2$. Notice
that the height of a cylindrical droplet can be expressed as,
\begin{equation}
  H=\sqrt{\frac{2V}{L_y}} \frac{1-\cos\theta_E}{\sqrt{2\theta_E-\sin 2\theta_E}},
\end{equation}
which is a monotonically increasing function of the contact angle. Thus the steady state velocity of the droplet
increases with its contact angle. As the surface becomes more hydrophobic, the shape of droplet becomes more like a
sphere, which reduces the viscous damping during the rolling motion. 
\begin{figure}[h!]
    \includegraphics[width=8cm]{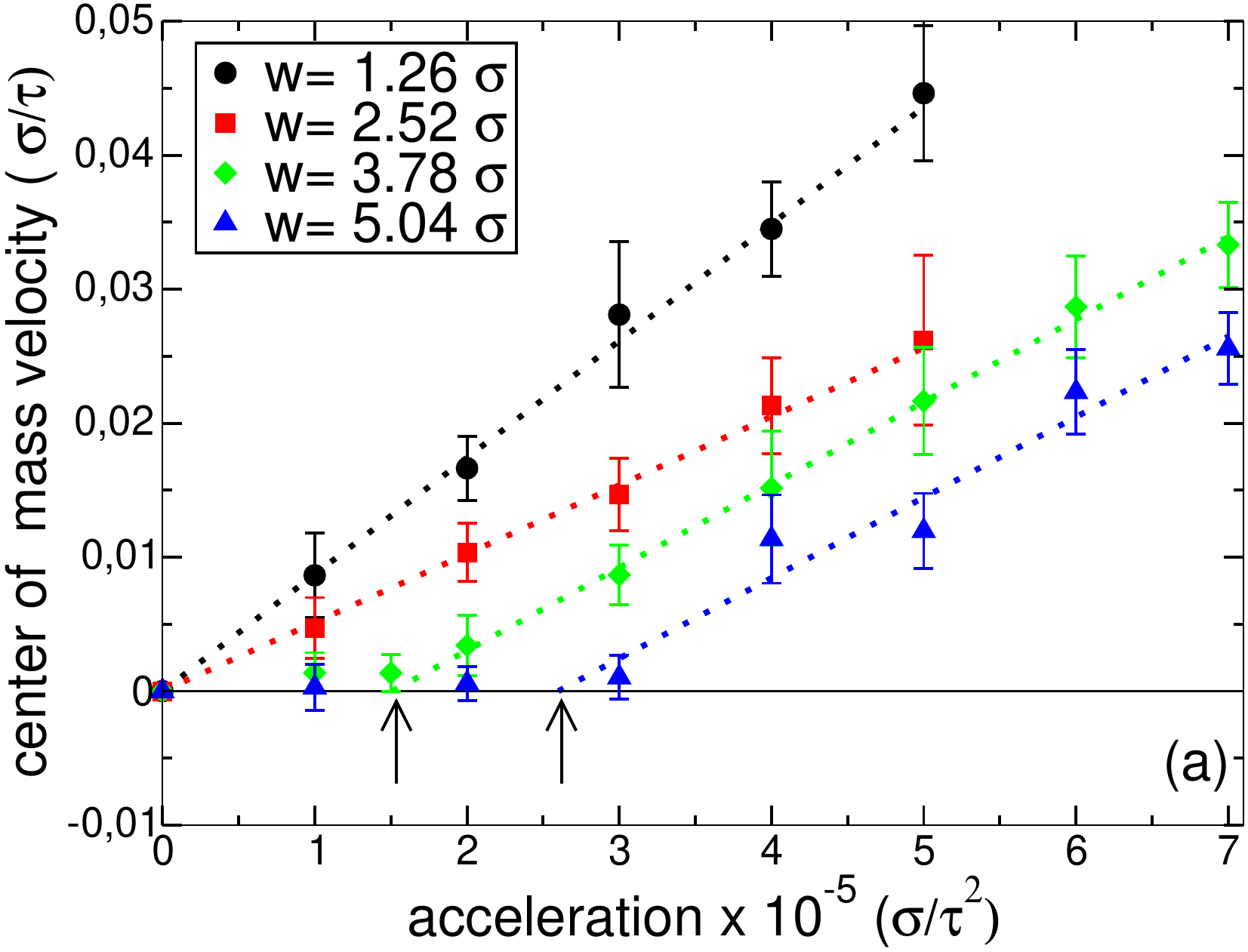} 
    \includegraphics[width=8cm]{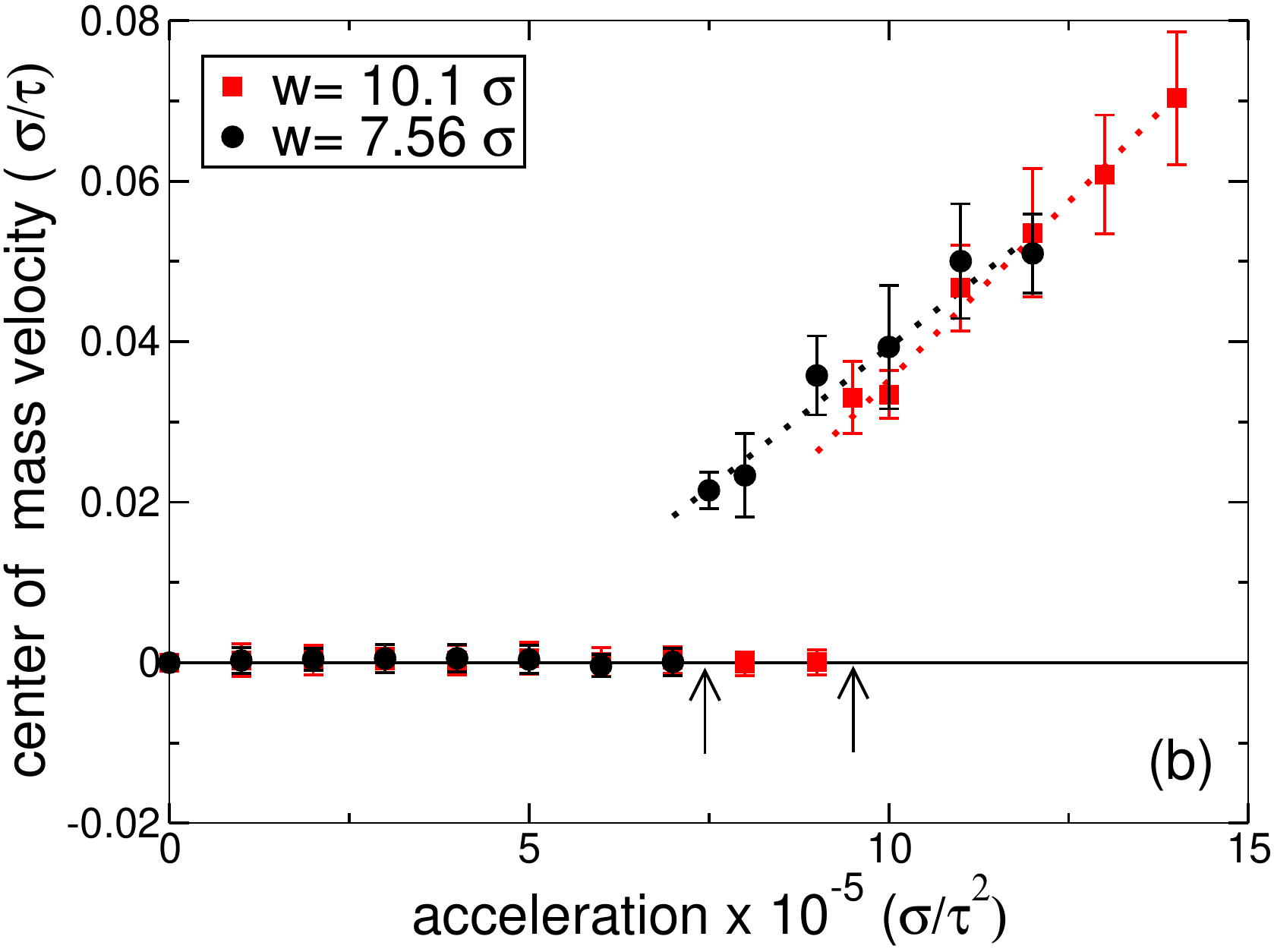}
    \caption{\label{fig:HETEva}
      (Color online) Velocity of the center of mass of the droplets for $N=50000$ monomers as a function of 
      the acceleration for the chemically heterogeneous surfaces.
      (a) $w=1.26-5.04\ \sigma$ and (b) $w=7.56 \sigma$ and $w=10.1\ \sigma$.
      The symbols correspond to the results of the MD simulations and the dotted lines to linear fits.
      The error bars correspond to the standard deviations of  the means. The arrows point to the
    depinning accelerations.}
\end{figure}

\begin{figure}[h!]  
 \includegraphics[scale=0.5]{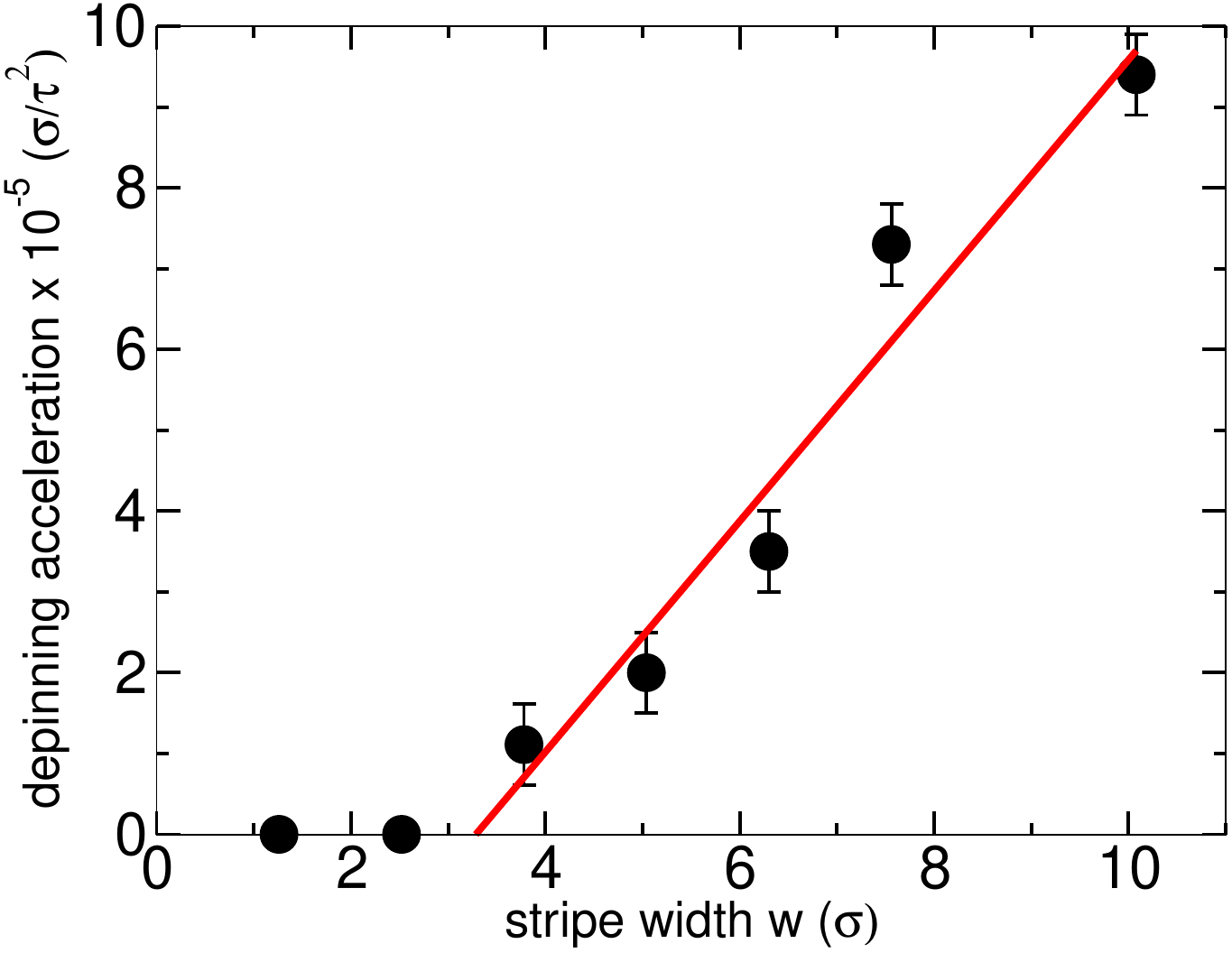}
 \caption{\label{fig:depinningivmeWp} The minimum depinning force as a function of the stripe width $w$. The circles
 are the results from the MD simulations and the solid line a linear fit on the non-zero values of the depinning acceleration.}
\end{figure}

\begin{figure}[h!]  
  \includegraphics[width=9cm,keepaspectratio=true]{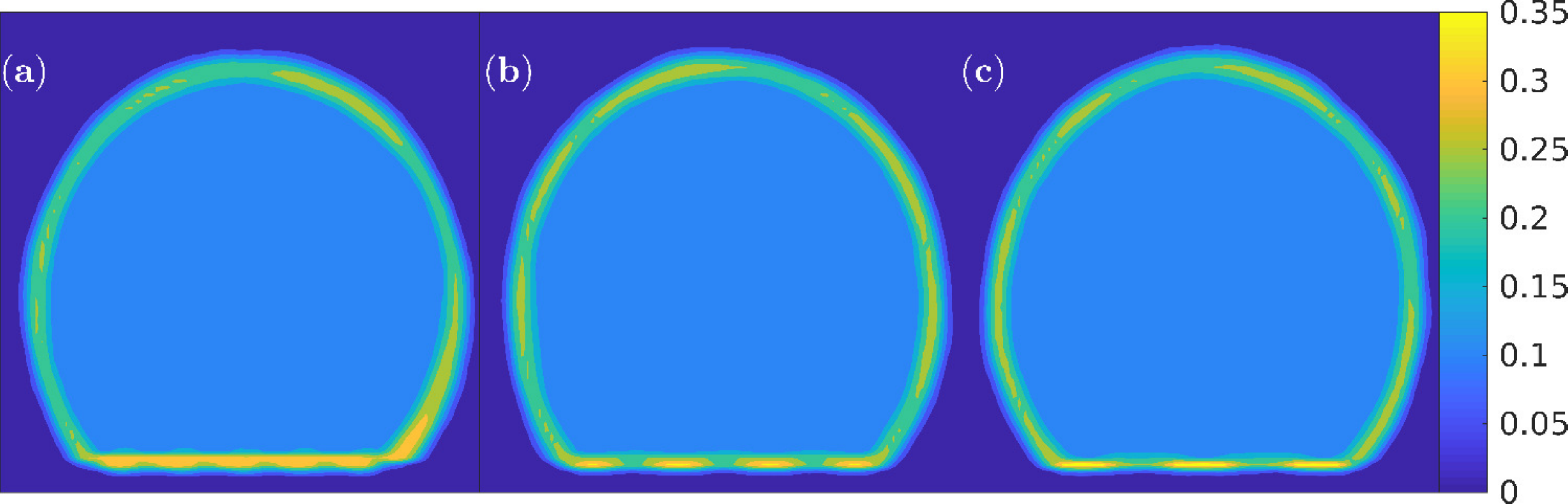}
  \caption{\label{fig:HETEcontour} (Color online) Density fluctuations of a pinned
    droplet with a bulk acceleration $a=0.00001\;(\sigma/\tau^{2})$. (a) $w=5.04\ \sigma$, (b) $w =7.56\ \sigma$ and
    (c) $w=10.1\ \sigma$. The contour maps correspond to the standard deviation 
    of the number density averaged over $2 \times 10^{6}$ simulation time steps. }
\end{figure}

\begin{figure}[h!]  
  \includegraphics[width=6cm]{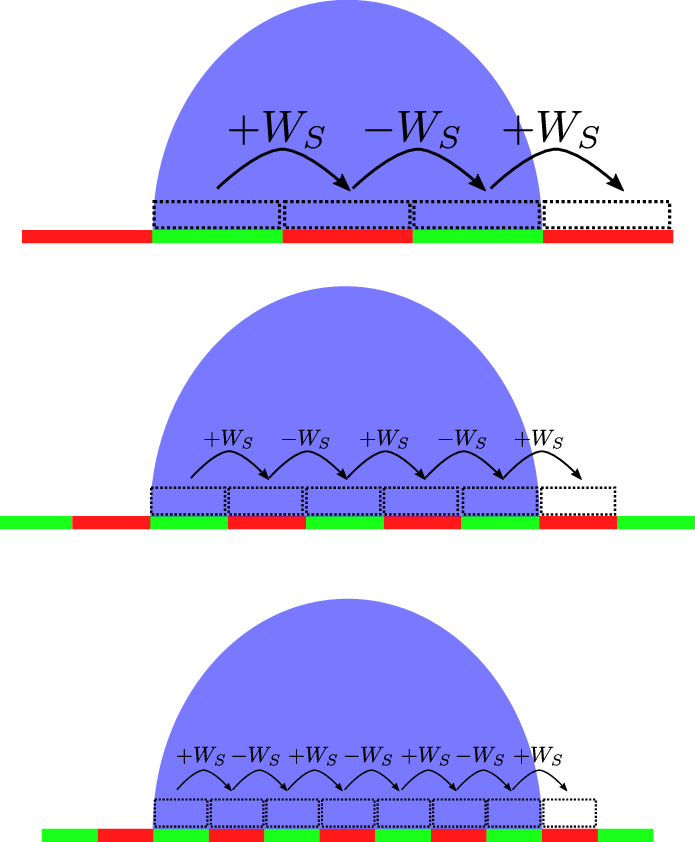}
  \caption{\label{fig:droplets} (Color online) An illustration for the depinning work $W_{s}$. A pinned droplet
    is in contact with $n$ hydrophobic stripes and $n+1$ hydrophilic stripes, the extremities must therefore be
  hydrophilic stripes.}
\end{figure}

We now focus on the chemically heterogeneous substrate. We consider the liquid droplets in
Fig. \ref{fig:4Width50000nsivi} and drive them with varying bulk
accelerations $a$.  Remark that we observed the contact angle hysteresis for $w_{p}\ge 23 \%$ values, 
the results for these systems are not given in this paper. We depict the time averaged velocity of
the center mass as a function of the acceleration for the different values of $w$ in Fig. \ref{fig:HETEva}.
For the two smallest stripe widths, namely $w=1.26\ \sigma$ and $w=2.52\ \sigma$ the velocity of the
center of mass is still a linear function of the acceleration. However, as the stripe width increase
we notice that a minimum acceleration is required for a sustained motion, in other words the droplet remains
pinned. Lets first focus on the pinned state. We notice that as the stripe width increases the minimum
acceleration increases. Using the results in Fig. \ref{fig:HETEva} we depict the critical acceleration
in Fig. \ref{fig:depinningivmeWp}. We notice that apart for very small stripe widths, the depinning acceleration
increases linearly with the stripe width. For small widths the energy barrier the droplet must cross is
relatively small, consequently thermal fluctuations are enough to overcome it.
On the other hand, the linear increase of the depinning force can be explained easily by a qualitative
argument. Indeed, when the droplet is pinned it maximizes its contact area with the hydrophilic stripes in order
to minimize the free energy. One can see this effect clearly from the density fluctuations in pinned
droplets in Fig. \ref{fig:HETEcontour}.  
In order to maximize its contact with the hydrophilic stripes, each extremity of the droplet has to be on a hydrophilic
stripe. In that case, if the droplet is in contact with $n$ hydrophobic stripes, it will be in contact with $n+1$
hydrophilic ones. Since there is slippage at the surface the whole contact area of the droplet moves
instead of only the contact line. Assuming that $+W_s$ is the work required to move the fluid of a hydrophilic stripe
to a hydrophobic one, then $-W_s$ is the work for the fluid moving from a hydrophobic stripe to a
hydrophilic. Since there is an extra hydrophilic stripe there will remain a net work $+W_s$ in order to depin
the droplet as schematized in Fig. \ref{fig:droplets}. 
The depinning work $W_s$ depends on the difference of surface energies $\Delta \gamma$ and on the
area of fluid to move, thus $W_s \sim w \Delta \gamma$. If the temperature is large enough,
$k_B T >W_s$, the thermal fluctuations are enough to depin the droplet. For lower temperatures, there will
be a critical stripe width after which the droplet is pinned, and the depinning work will increase
linearly with $w$.

\begin{figure}[h!]  
\includegraphics[width=4.25cm]{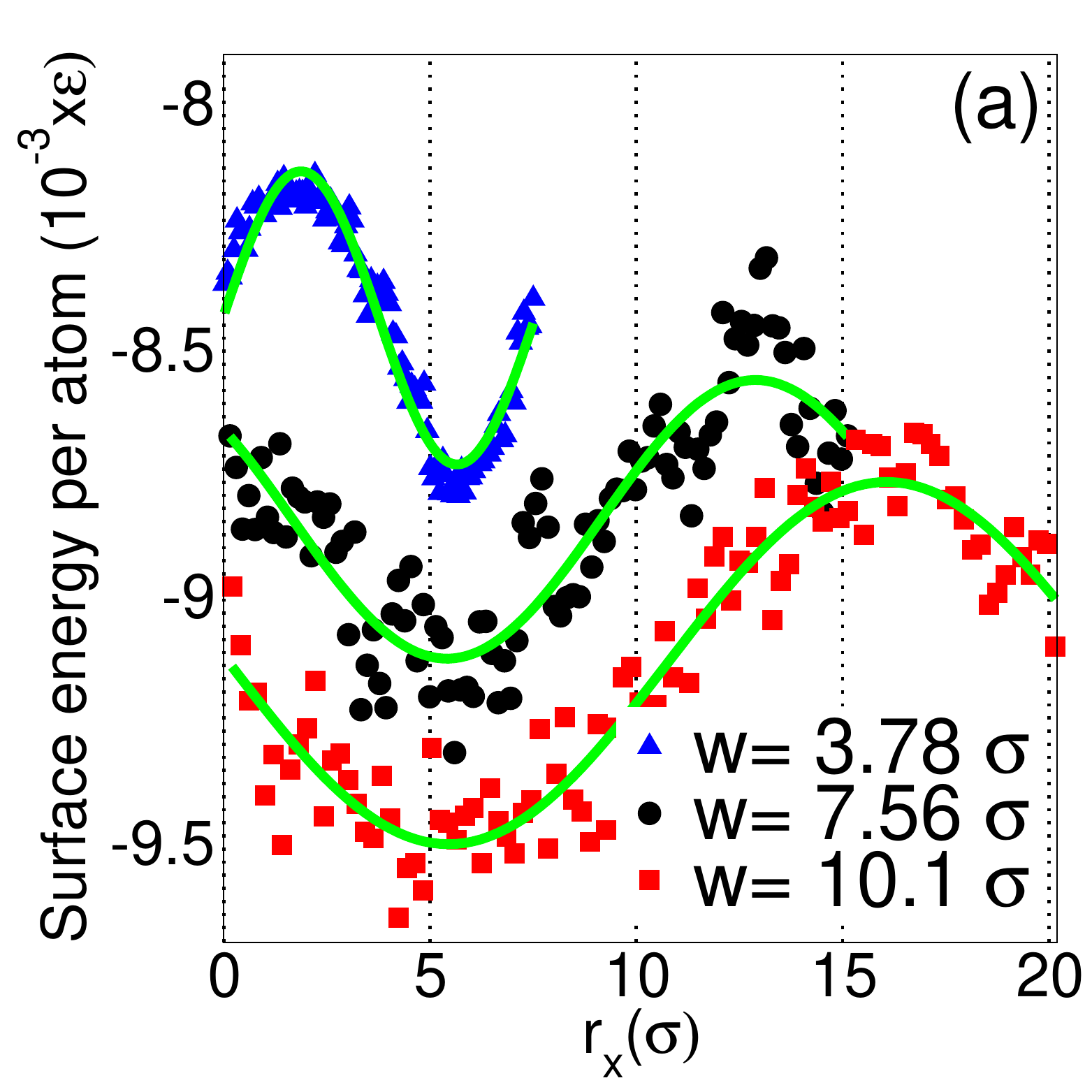}
 \includegraphics[width=4.25cm]{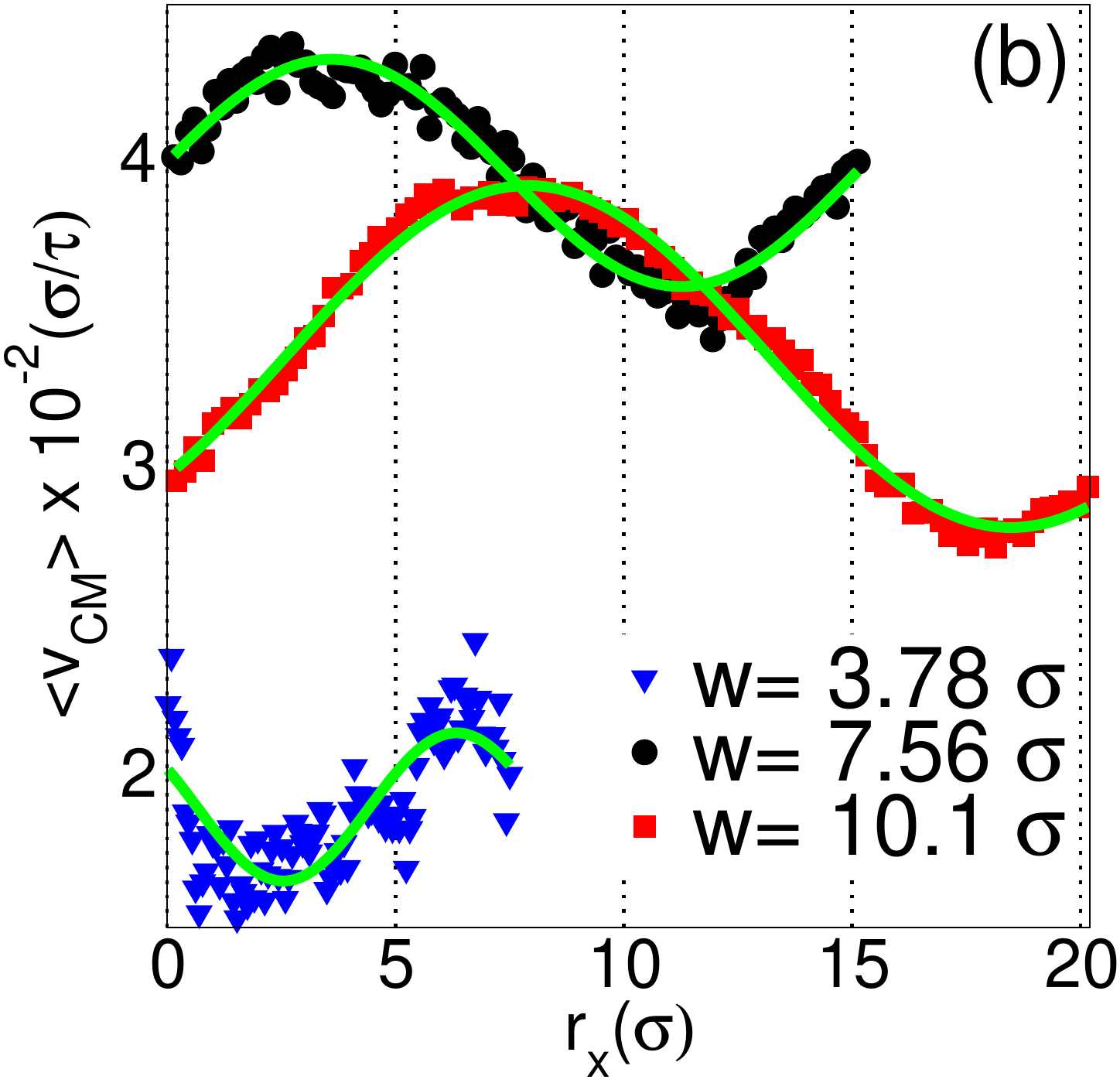}
 \includegraphics[width=6cm]{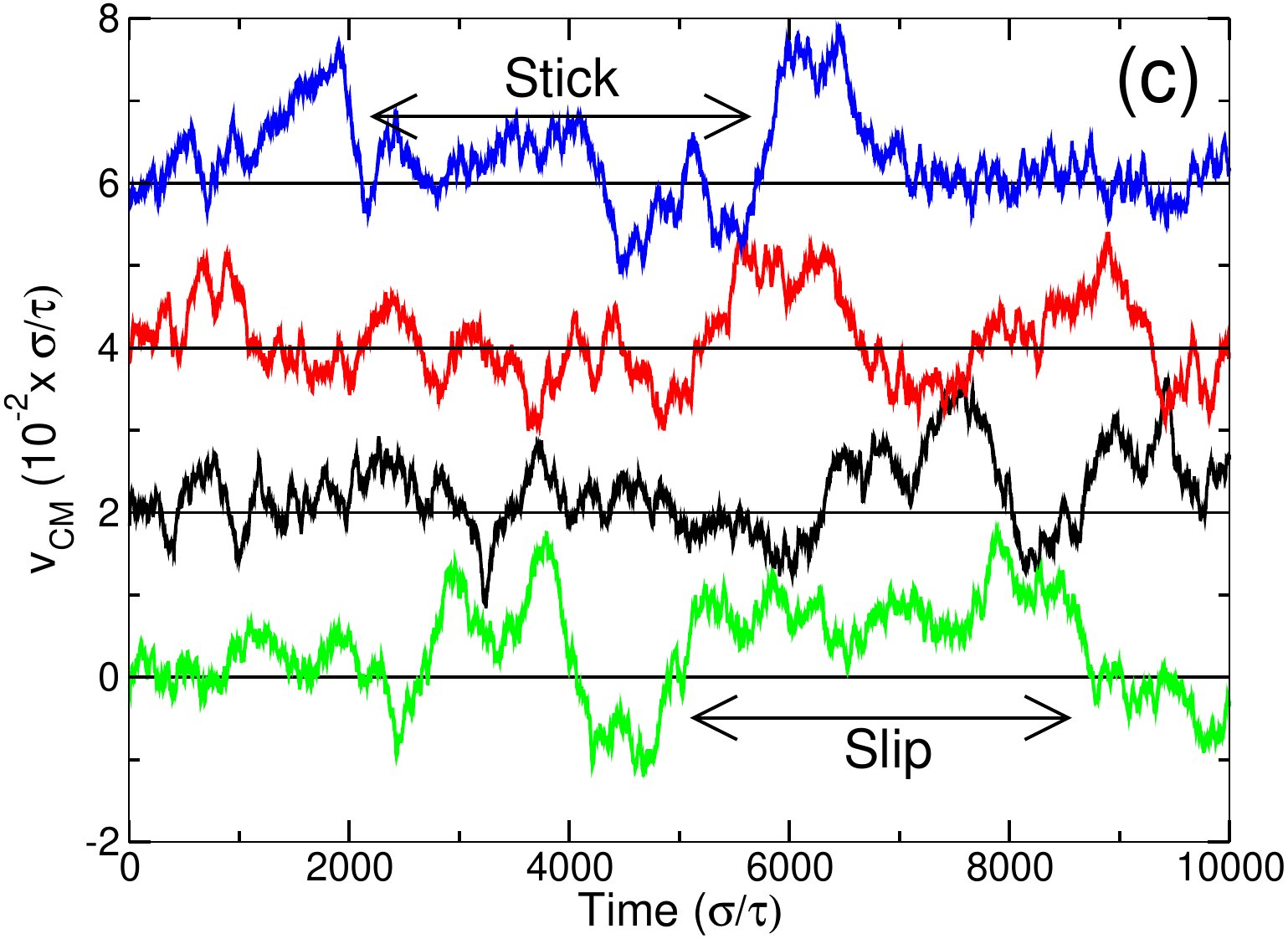}
 \caption{\label{fig:W20UsFsVCMx}(Color online) (a) The surface energy per atom and (b) The averaged velocity
   of the center of mass as a function of the position of the center of mass. The plain lines correspond
   to sinusoidal fits. The center of mass velocity for $w=3.78 \sigma$ is multiplied by 5 for better visualization.
   (c) Velocity of the center of mass as a function of time for four typical trajectories in the stick-slip regime, the
 velocity is translated.}
\end{figure}

We now focus on the dynamics of the droplet after the depinning. For the two smallest stripe width, we do not observe
any pinning, and $\mean{v_{\rm CM}}$ increases linearly with the acceleration. Assuming the model for the homogeneous
substrates is still valid, we perform linear fits on $\mean{v_{\rm CM}}$ and evaluate the effective damping
coefficients. We give in Table \ref{tab:HeteDynamic} the results of the simulations and the value of $\gamma$ obtained
from the slip lengths calculations and geometry of the droplet with Eqs. \ref{eqmodel1} and \ref{eqmodel3}. Again we
notice that the dynamics is relatively well described in terms of the model. For small stripe widths, the fact that
the substrate is chemically heterogeneous does not alter significantly the dynamics. Except for a smaller
slip length $\delta$, and thus a larger effective damping coefficient. This is due to the fact that while the surface
is heterogeneous it is still smooth, the absence of surface roughness permits to slide and rotate easily on the
substrate without any significant changes to the velocity profile inside the droplet.
\begin{table}[h!]
  \caption{\label{tab:HeteDynamic}Molecular Dynamics simulation 
    results for polymer drops of $N=50000$ atoms on chemically 
    heterogeneous surfaces for different stripe widths.}
  \begin{ruledtabular}	
    \begin{tabular}{cccccc}
      $w (\sigma)$& $r_z\ (\sigma)$ & $H\ (\sigma)$ & $\delta\ (\sigma)$ & $ \gamma \; (m/\tau) $ & $ \gamma_{\rm MD} \; (m/\tau) $\\
      \hline
      1.26   & 27 & 59 & 74 & 60     &     56 \\
      2.52   & 27 & 58 & 64 & 68     &     96 \\
      3.78   & 27 & 59 & 62.5 & 68   &     81 \\
      5.04  & 26 & 57 & 62.5 & 71    &     83\\
      7.56  & 26 & 57 & 62.5 & 71    &     71 \\
      10.08  & 27 & 58 & 62.5  & 71  &     57 \\
    \end{tabular}
  \end{ruledtabular}		
\end{table}
Similarly, for the pinned droplets, after the critical acceleration, for large enough accelerations, we recover
a linear regime for the velocity of the center of mass. We observe a sinusoidal variation of the velocity of
the droplet as a function of time. As the edge of the droplet crosses from a hydrophilic stripe to a hydrophobic one,
the total surface energy increases since the interaction is less attractive, and consequently, the kinetic energy
increases. We have evaluated the average velocity and surface potential energy as a function of the position
of the center of mass with respect to the substrate. We averaged the results over the distance of two stripes for
better statistics, the results are depicted in Fig. \ref{fig:W20UsFsVCMx}(a) and Fig. \ref{fig:W20UsFsVCMx}(b) for the two
largest stripes at an acceleration of $a=10^{-4}\ \sigma/\tau^{2}$ and for $w=3.78\ \sigma$ at $a=1.5\times 10^{-5}\ \sigma/\tau^2$.
When the surface energy is minimum, the center of mass of the droplet corresponds to the pinned position, in other
words when it is in contact with one more hydrophilic stripe than a hydrophobic one. As the droplet crosses to a hydrophobic
stripe its velocity decreases until it is in contact with one more hydrophobic stripe than a hydrophilic one, corresponding to the
largest potential energy. Afterwards the velocity increases again. We notice that the modulation of
the center of mass velocity does not affect significantly its mean value. Indeed, the linear fits in
Fig. \ref{fig:HETEva} and the corresponding results in Table \ref{tab:HeteDynamic} suggest that after the critical
acceleration the dynamics of the droplet can still be relatively accurately described by the model.

On the other hand, one would expect that for accelerations just above the critical acceleration the dynamics
would be consistent with stick-slip motion, in other words a succession of pinnings and depinnings. In that case,
for a droplet in the pinned state, large thermal fluctuations allows to depin the droplet until it crosses to the
next stripe and is pinned again. We do observe this behavior on individual trajectories as depicted in Fig.
\ref{fig:W20UsFsVCMx}(c), however the time between depinnings varies wildly. Hence, once we consider time averages
or the average on different trajectories one recovers the sinusoidal variation of the velocity as for the higher
accelerations.

 \section{Conclusion}
 In this paper we presented coarsed-grained molecular dynamics simulation of the statics
 and dynamics of cylindrical polymer droplets on chemically homogeneous and heterogeneous
 surfaces. The surfaces consist of two layers of {\it fcc} lattices which interact with a modified
 Lennard-Jones potential with the polymeric fluid. The hydrophobicity of the surfaces is tuned with an
 empirical parameter weighting the attractive term. Chemically heterogeneous surfaces can then be
 defined with stripes of different hydrophobicity.  We first evaluated the equilibrium
 contact angle on different surfaces. We showed that at equilibrium the droplet deforms slightly
 in order to accommodate one extra hydrophilic stripe with respect to hydrophobic
 ones. As a result, at equilibrium, the droplet has to be in contact with an odd number of
 stripes. As the stripe width increases this results in relatively large differences of contact
 angle. However, on average we have observed that the Cassie-Baxter relation gives a good
 approximation of the equilibrium contact angle.

 We then focused on the boundary condition, indeed at microscopic scales the fluid can slip on
 the solid surface. This results in a combination of sliding and rotating motion for small
 droplets. We previously showed that on homogeneous surfaces, the steady-state velocity of the droplet
 scales linearly with the acceleration and depends only on its geometry i.e. contact angle and size,
 and the amount of slippage at the surface.  \cite{ServantieMuller_JCP128_2008} For small stripe widths, this
 is still true as the fluid only sees an effective surface, however, as the stripe width increases we noticed
 that the droplet becomes pinned until a sufficiently large acceleration is exerted. We showed that the depinning
 acceleration increases linearly with the stripe width. Since at equilibrium, the droplets extremities have to be
 on hydrophilic stripes, the net work required to depin the droplet is the work to move an amount of fluid from a
 hydrophilic stripe to an hydrophobic one, which in turn scales as the surface of the stripes. Once the droplet
 is depinned the steady-state velocity oscillates with time, consistent with the changes in surface energy when
 crossing different stripes. Finally, between the pinned state and linear regime we observed a characteristic
 stick-slip regime where large thermal fluctuations can depin the droplets.

\acknowledgements
This research is financially supported by {\.I}T{\"U} BAP under Grant No. 38062.

\bibliographystyle{aipauth4-1} 

%

\end{document}